\documentclass[final,3p, twocolumn]{elsarticle}
\usepackage{amssymb}
\usepackage{lipsum}
\usepackage{float}
\usepackage{comment}
\usepackage{amsmath,amsthm,amsfonts,bm,epsfig,color,empheq,graphicx,graphics,pgfplots}
\usepackage{framed} % Framing content
\usepackage{multicol} % Multiple columns environment
\usepackage{caption}
\usepackage{subcaption}
\usepackage{lineno}

\usepackage{tikz}
\usetikzlibrary{shapes.geometric, arrows.meta, positioning, fit, backgrounds}

\tikzstyle{startstop} = [rectangle, rounded corners, minimum width=2.5cm, minimum height=1cm, text centered, draw=black, fill=white]
\tikzstyle{process} = [rectangle, minimum width=2.5cm, minimum height=1cm, text centered, draw=black, fill=white]
\tikzstyle{arrow} = [thick,->,>=Stealth]
\journal{Results in Physics}

\begin{document}

\begin{frontmatter}

%% Title, authors and addresses

%% use the tnoteref command within \title for footnotes;
%% use the tnotetext command for theassociated footnote;
%% use the fnref command within \author or \affiliation for footnotes;
%% use the fntext command for theassociated footnote;
%% use the corref command within \author for corresponding author footnotes;
%% use the cortext command for theassociated footnote;
%% use the ead command for the email address,
%% and the form \ead[url] for the home page:
%% \title{Title\tnoteref{label1}}
%% \tnotetext[label1]{}
%% \author{Name\corref{cor1}\fnref{label2}}
%% \ead{email address}
%% \ead[url]{home page}
%% \fntext[label2]{}
%% \cortext[cor1]{}
%% \affiliation{organization={},
%%            addressline={}, 
%%            city={},
%%            postcode={}, 
%%            state={},
%%            country={}}
%% \fntext[label3]{}

\title{A Data-Driven Model for the Field Emission from Broad-Area Electrodes}

%% use optional labels to link authors explicitly to addresses:
%% \author[label1,label2]{}
%% \affiliation[label1]{organization={},
%%             addressline={},
%%             city={},
%%             postcode={},
%%             state={},
%%             country={}}
%%
%% \affiliation[label2]{organization={},
%%             addressline={},
%%             city={},
%%             postcode={},
%%             state={},
%%             country={}}

\author[]{Moein Borghei}
\ead{mborghei@avalanchefusion.com}
\author[]{Robin Langtry}

\affiliation{organization={Avalanche Energy},%Department and Organization
            addressline={9100 E Marginal Way S}, 
            city={Tukwila},
            postcode={98108}, 
            state={WA},
            country={USA}}

\begin{abstract}
Electron emission from cathodes in high field gradients is a quantum tunneling effect. The 1928 Fowler-Nordheim field emission (FE) equation and the 1956 Murphy-Good FE equation have traditionally been key in describing cold field emissions, offering estimates for emitters for almost a century. Nevertheless, applying FE theory in practice is often constrained by the lack of data on the distribution and geometry of the emission sites. Predictions become more challenging with an uneven electric field distribution at the cathode surface. Consequently, FE formulations are frequently calibrated using current-voltage data after test, limiting their efficacy as true predictive models.

This study develops an alternative model for field emission using a data-driven predictive approach based on (1) vast experimental data, (2) electrostatic simulations of the cathode surface, and (3) detailed material and geometry properties, which together overcome these limitations. The objective of this work is to develop and harness this comprehensive dataset to train a machine learning model capable of providing precise predictions of the cathode current in order to further the understanding and application of field emission phenomena. More than 259 hours of experimental data have been processed to train and benchmark some of the well-known machine learning models. After two stages of optimization, a coefficient of determination $>98\%$ is achieved in the prediction total field emission current using ensemble models.
\end{abstract}

%%Graphical abstract
%\begin{graphicalabstract}
%\includegraphics{grabs}
%\end{graphicalabstract}

%%Research highlights
%\begin{highlights}
%\item Research highlight 1
%\item Research highlight 2
%\end{highlights}

\begin{keyword}
%% keywords here, in the form: keyword \sep keyword, up to a maximum of 6 keywords
field emission \sep Fowler-Nordheim theory \sep machine learning \sep quantum tunneling  

%% PACS codes here, in the form: \PACS code \sep code

%% MSC codes here, in the form: \MSC code \sep code
%% or \MSC[2008] code \sep code (2000 is the default)

\end{keyword}

\end{frontmatter}

%\tableofcontents

%% \linenumbers

%% main text

\section{Introduction}\label{sec:intro}
Conductors are often used to transport electrical charges. In most applications, these electrons are totally confined within the material by the presence of a surface potential barrier known as the material's work function. Until the early twentieth century, the emission of electrons from conductors was engineered through thermionic and photoelectric emission processes. 

Interest in field emission arose in the 1910s through J. E. Lillienfield's work developing portable X-ray machines \cite{Lilienfeld_1922}, and later through W. D. Coolidge's observation of a new phenomenon of electron emission from the cold cathode due to the high electric field strength at the surface \cite{coolidge_1928}. Following a series of experimental and theoretical works \cite{schottky_uber_1923, millikan_1926, millikan_1928, richardson_1928, Oppenheimer_1928_true, Lauritsen1929}, Fowler and Nordheim derived a closed-form equation for the local emission current density of an emission site in 1928 \cite{fowler_1928}. Based on the 1D Schrödinger equation, Fowler-Nordheim derived an analytic formulation to approximate the tunneling rate of electrons through the potential barrier as a function of the local electric field and the work function of the material. Stern found and corrected a numerical error in the Fowler-Nordheim equation, and also introduced the concept of what is now called the Fowler-Nordheim plot \cite{Stern_1929}.

%Field emission is a phenomenon characterized by the emission of electrons from a solid surface under the influence of a strong electric field. 

%Since then, field emission has been found to have applications in electron microscopy \cite{pawley1997}, X-ray sources \cite{yue2002}, flat panel displays \cite{talin2001}, electron beam lithography \cite{stephani1983}, electron gun \cite{crewe1968}, microwave amplifiers \cite{whaley2000}, and more. Nevertheless, field emission has also presented challenges to maintaining high voltage, as the precursor to vacuum breakdown \cite{latham1983} and a power loss term that affects system efficiency. 

%Serious errors were subsequently found in the 1920s work on field emission. The outcome was the publication by Murphy and Good (MG) in 1956 [**] of a corrected FE equation that predicts much higher local emission current densities. Equation (1) below is the zero-temperature version of this 1956 MG FE equation. In FE equations, the field enhancement factor …"

\begin{figure}
    \centering
    \includegraphics[width=0.35\textwidth]{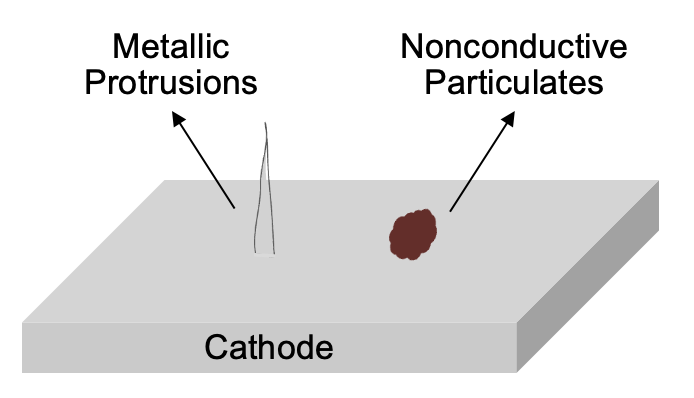}
    \caption{Exemplary illustration of the field enhancement features. }
    \label{fig:field-enhancement}
\end{figure}

Notable errors were found in the 1920s analytic solution of Fowler-Nordheim. These were corrected in 1956 by Murphy and Good \cite{murphy_good_1956} who proposed a corrected FE equation predicting considerably higher local current emission densities. Since then, the local electric field of the emitter is explained by field enhancement factors. Traditionally, sharp metallic whiskers that can be created artificially or developed through surface dislocation activity \cite{engelberg_2018} or surface diffusion \cite{Jansson_2020} under high field strength were thought to cause enhancement (see Figure \ref{fig:field-enhancement}). Over the years, efforts have been made to approximate the shape of field enhancing features \cite{zuber_2002, Edgcombe_2005, Edgcombe_2007, Kyritsakis_2015}. Despite these efforts, the observation of field emission current on an extremely polished cathode surface indicates alternative sources of field emission, apart from whiskers. This has been confirmed in \cite{Cox_1975}. Additionally, an observed shift in electron emission spectra to energies below the Fermi level \cite{ATHWAL_1981, Athwal_1984} and the cathode electroluminescence \cite{Hurley_1979}, further point towards a nonmetallic origin.

Aside from the debates over the origin of the field emission current, nearly a century later after the discovery of the field emission, there is still no widely accepted model for the prediction of total field emission for practical applications involving broad-area electrodes, large gaps, and nonuniform electric field distribution where numerous emission sites make up the total field emission current, or \emph{dark current}. In practice, it is nearly impossible to acquire all the geometric details of the microscopic emission sites on a broad-area electrode.

In this work, our aim is to train a data-driven model based on more than 259 hours of experimental data. We supplement the input dataset with electrostatic, geometric, and material information about the cathode. The model is agnostic to the individual field emission sites and rather aims to recognize their collective pattern on a broad-area electrode. To complement the cathode voltage and current measurements, we utilize the electric field distribution from an electrostatic simulation, which yields statistical features that explain the probability distribution of the electric field magnitude. Optical microscopy of a $1 \ \mu m \times 1 \ \mu m$ section of a cathode data provides statistical features of the projection heights of the surface structure. Lastly, the work function of the material and the total cathode area conclude the input dataset. 

Relaxing the modeling of broad-area emission from the traditional curve fitting to a more data-informed complex model allows one to capture a wider variation of data. Moreover, such a model can be constantly reinforced with more test results, provided that the aforementioned set of data is provided. Performance comparison of different supervised machine learning models on the dataset shows that we can achieve $>98\%$ accuracy in the prediction of the total field emission current with the gradient boosting ensemble model. 

The structure of the remainder of this article is delineated as follows: In Section \ref{sec:FN}, we provide a brief review of the Murphy-Good theory of field emission with a focus on key aspects of the theory relevant to our work. Section \ref{sec:framework} elucidates the framework of our approach to field emission prediction, which will be discussed in more detail in subsequent sections. The experimental setup is briefly discussed in Section \ref{sec:exp}. Section \ref{sec:data} details the preprocessing of data for application in machine learning models. The results of training and validation of various machine learning models are presented in Section \ref{sec:results}. Finally, a discussion of the results and the conclusions are presented in Sections \ref{sec:discuss} and \ref{sec:conclude}, respectively.

\section{Overview of Murphy-Good Field Emission Theory}\label{sec:FN}
Murphy and Good \cite{murphy_good_1956} used quantum tunneling theory to calculate the probability that an electron escapes from a metal surface under a high electric field, which is then integrated into the energy states to obtain the local emission current density. The zero-temperature version of the Murphy-Good equation for steady electric fields is as follows:

\begin{equation}
    J_L = \frac{a \ E^2}{\phi \ {t^2(y)}} \ e^{\displaystyle -\frac{b \ \phi^{3/2} \  v(y)}{E} }
\end{equation}

where $a$ and $b$ are constants, $E$ is the intensity of the local electric field at the emitter, and $\phi$ is the work function of the material. Finally, $t(y)$ and $v(y)$ are slowly varying functions of $\phi$ and $E$. $t(y)$ is close to unity, but $v(y)$ takes values slightly less than unity. 

The deviation of $v(y)$ from unity is responsible for the large difference between the predictions of the Fowler-Nordheim and the Good-Murphy equation. Forbes and Deane provide a simple yet accurate approximation for $v(y)$ in \cite{forbes_2007}, and Miller reports the tabulated data in \cite{miller_1980}. For additional review of recent contributions to the Nordheim parameter and more, see \cite{Forbes2020, jensen_2024}.

The local electric field at the emission site, denoted $E$, frequently exceeds the expected magnitude derived from the estimation of the macroscopic field $E_0$, by two to three orders of magnitude. Consequently, an effective enhancement factor, $\beta_{eff} = \frac{1}{I} \int_{S} \beta(\textbf{x}) J(\textbf{x}) dS$, is typically used to approximate the microscopic electric field \cite{kirkpatrick_1992}. 
However, using $E = \beta_{eff} E_0$ usually does not yield accurate results. A more practical approach is to use the dimensionless field enhancement factor at its apex $\gamma_a$, which defines the local electrostatic field as $E = \gamma_a E_0$. - assuming a cylindrically symmetric protrusion \cite{deassis2019}.  

To derive the total emission current from an emission site, the local emission current density can be associated with the total emission current via the relation $I = A J_L$, where $A$ represents the notional emission area of the emitter which is often field dependent \cite{forbes_2007}.

%All these complication and craps gets even harder for broad-area electrodes, that's why here we propose a method based on data-driven models to incorporare collective behavior of the field emission current. 

%In practical cases, only after acquiring the $I-V$ measurement can the field enhancement factor be derived from the linear slope of $log(I/E^2)$ versus $1/E$. For many, this has been the confirmation of the famous whisker model that originated in the mid-60s that the applied macroscopic field is geometrically enhanced by a factor at the tip of a whisker to reach the required threshold and the microscopic electric field exceeds the threshold value of $3 \ GV/m$. As mentioned in Section \ref{sec:intro}, some experimental results have shown that the emission came from a more complex regime involving an insulating microinclusion as shown in Figure \ref{fig:field-enhancement}.

%If one knows the emitting area of an emission site, $A$, the total current from the emission site, $I$, can be derived by integrating the local emission current density into the emitting area. For computational purposes, the total current can be taken as the product of the apex current density $J$ with a notional emission area, and that area changes approximately linearly with the apex field due to the way the current density decreases.

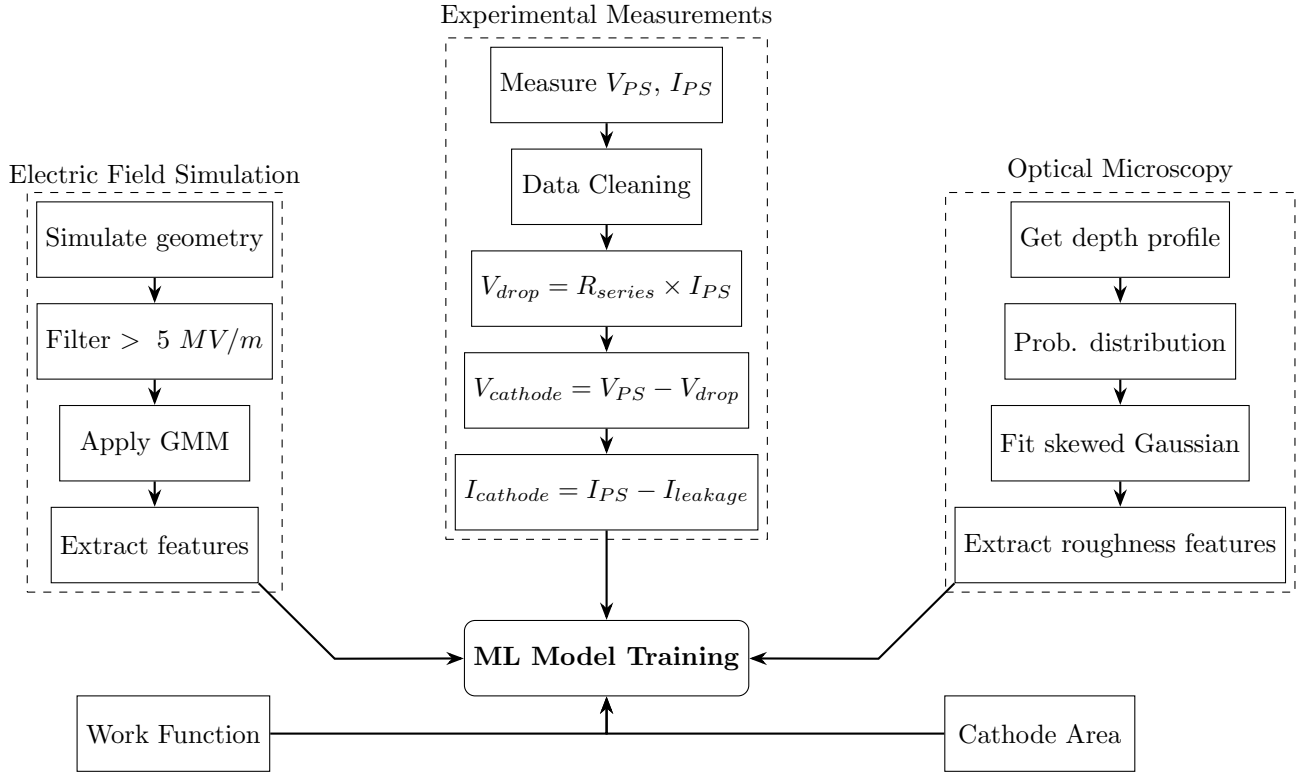
\begin{figure*}[h!]
\centering
\begin{tikzpicture}[node distance=1.2cm and 1.2cm, auto, every node/.style={align=center}]

% Central goal node
\node (goal) [startstop] {\textbf{ML Model Training}};

% Experimental Measurements group
\node (exp_meas) [process, above of=goal, yshift=1cm] {$I_{cathode}=I_{PS}-I_{leakage}$};
\node (exp1) [process, above of=exp_meas, yshift=0.15cm] {$V_{cathode} = V_{PS}-V_{drop}$};
\node (exp2) [process, above of=exp1, yshift=0.15cm] {$V_{drop}=R_{series}\times I_{PS}$};
\node (exp3) [process, above of=exp2, yshift=0.15cm] {Data Cleaning};
\node (exp4) [process, above of=exp3, yshift=0.15cm] {Measure $V_{PS}$, $I_{PS}$};

% Electric Field Simulation group
\node (sim4) [process, left=of goal, xshift=-1.5cm, yshift=1.5cm] {Extract features};
\node (sim3) [process, above of=sim4, yshift=0.15cm] {Apply GMM};
\node (sim2) [process, above of=sim3, yshift=0.15cm] {Filter $> \ 5 \ MV/m$};
\node (sim1) [process, above of=sim2, yshift=0.15cm] {Simulate geometry};

% Electron Microscopy group
\node (mic4) [process, right=of goal, xshift=1.5cm, yshift=1.5cm] {Extract roughness features};
\node (mic3) [process, above of=mic4, yshift=0.15cm] {Fit skewed Gaussian};
\node (mic2) [process, above of=mic3, yshift=0.15cm] {Prob. distribution};
\node (mic1) [process, above of=mic2, yshift=0.15cm] {Get depth profile};

% Other groups
\node (work_func) [process, left of=goal, yshift=-1cm, xshift=-4.5cm] {Work Function};
\node (cath_area) [process, right of=goal, yshift=-1cm, xshift=4.5cm] {Cathode Area};

% Ellipses around groups
\begin{scope}[on background layer]
    \node[draw, dashed, rectangle, fit={(sim1) (sim4)}, label=above:Electric Field Simulation] {};
    \node[draw, dashed, rectangle, fit={(exp_meas) (exp4)}, label=above:Experimental Measurements] {};
    \node[draw, dashed, rectangle, fit={(mic1) (mic4)}, label=above:Optical Microscopy] {};
\end{scope}

% Draw arrows to the central goal
\draw [arrow] (exp_meas) -- (goal);
\draw [arrow] (sim4.south east) -- ++(1,-1) -- (goal);
\draw [arrow] (mic4.south west) -- ++(-1,-1) -- (goal);
\draw [arrow] (work_func.east) -- ++(1,0) -| (goal);
\draw [arrow] (cath_area.west) -- ++(-1,0) -| (goal);

% Draw arrows for sub-steps
\draw [arrow] (exp4) -- (exp3);
\draw [arrow] (exp3) -- (exp2);
\draw [arrow] (exp2) -- (exp1);
\draw [arrow] (exp1) -- (exp_meas);

\draw [arrow] (sim1) -- (sim2);
\draw [arrow] (sim2) -- (sim3);
\draw [arrow] (sim3) -- (sim4);

\draw [arrow] (mic1) -- (mic2);
\draw [arrow] (mic2) -- (mic3);
\draw [arrow] (mic3) -- (mic4);

\end{tikzpicture}
\caption{Flowchart illustrating the feature extraction and machine learning model training process for field emission.}
\label{fig:flowchart}
\end{figure*}

\subsection{From Single Emitter to Multiple Emitters}
In practical applications, the measured field emission current includes the contribution of not one but many emitters. Ideally, the field enhancement factor, work function, and emission area would be known for each emission site. However, it is merely impossible to do so except for the case of electrodes which have been specifically engineered for a particular enhancement factor. 

Tomaschke and Alpert \cite{tomaschke_1967} examined the collective field emission of 100 emission sites, where each site is randomly assigned a field enhancement factor and an emission area. The total accumulation of the emitters' contributions to the total current is then given as:

\begin{equation}
    I = \sum_{i=1}^{N} A_i \frac{a \ {E_i}^2}{{\phi_i} \ {t(y)}^2} \ e^{\displaystyle -\frac{b \ {\phi_i}^{3/2} \  v(y)}{E_i} }
\end{equation}

Their findings reveal that the total emission is closely aligned with a linear Fowler-Nordheim plot. Therefore, they can determine an effective field enhancement factor and a total field-emitting area. Several studies \cite{feng_2005,farrall_1970} have sought to find an effective field enhancement factor for broad-area cathodes. The common observation of these studies is that as the applied electric field increases, $\beta_{eff}$ decreases. As the electric field increases, the lower $\beta_{eff}$ areas on the surface begin to contribute more to the local current density.

The inability of these models to adapt to more complex geometries and to provide a systematic approach to predict the field emission current from practical electrode geometries necessitates the development of a novel model based on experimental data, agnostic to the individual emitter characteristics while correctly capturing the collective behavior of broad-area electrodes in terms of field electron emission.

\section{Predictive Modeling of Field Emission Current Using Data-Driven Approaches} \label{sec:framework}

Based on the theoretical overview of field emission, the key parameters of field emission are classified into these categories: (a) electrostatic parameters that describe the macroscopic field gradient at the cathode surface, (b) geometric parameters that influence the microscopic field gradient at the cathode surface and the emitting area, and (c) the chemical composition of the cathode surface. This study assumes that the vacuum level and the residual gas composition remain reasonably constant across different tests.

In this study, the data collected are processed and utilized to train a machine learning model to predict the field emission current of the cathode based on a given voltage. The objective is to transition from curve fitting the Fowler-Nordheim equation to a generalized, nonlinear model that more accurately captures the relationship between cathode voltage and current, and that better incorporates the described statistical features. This framework is depicted in Figure \ref{fig:flowchart}.

At the center of this process is machine learning (ML) model training, which integrates input from three primary groups: experimental measurements, electric field simulation, and optical microscopy. Additionally, factors such as work function and cathode area are considered, ensuring a holistic approach to training.

Starting with Experimental Measurements, the process begins by measuring the current-voltage ($I-V$) characteristics. This involves calculating the voltage drop across the series resistor and subsequently adjusting the measured voltage and current to obtain the actual cathode voltage, $V_{cathode}$ and current, $I_{cathode}$.  $V_{cathode}$ is used as one of the input parameters and $I_{cathode}$ is the target quantity for the approximation.

In parallel, Electric Field Simulation involves simulating the geometry explained in Section \ref{sec:exp}, filtering fields above $5 \ MV/m$ to focus on significant emission areas, applying a Gaussian mixture model (GMM) for statistical analysis, and extracting relevant features that make up the input matrix.

Optical microscopy is used to collect detailed surface information. This involves obtaining depth profiles, determining probability distributions, fitting skewed Gaussian models, and extracting roughness features. Although the emission area and field enhancement factor are not explicitly provided to the data-driven models, the statistical features derived from microscopy images supply the model with information about the density and profile of emitters. The machine learning models capture the variations in the emission area with the electric field through correlations in the provided data.

Lastly, two features are added to complete the input features vector. One is the cathode work function, which affects electron emission properties, and the second is the total cathode area which is used to extrapolate the emitter distribution from the localized microscopy images to the whole surface. Together, these diverse inputs ensure that the ML model is trained on a robust dataset, which includes empirical measurements, theoretical simulations, and microscopic surface characteristics, allowing accurate and reliable predictions of field emission behavior.

%To make this model operational, it is necessary to have parameters such as the electric field, surface roughness, work function, and cathode area.
%When a collection of physical and environmental characteristics is used as input features, $X$ to forecast the total field emission current, $y$, the task is identified as a regression analysis.

\section{Experiment Setup}\label{sec:exp}
\begin{figure}
    \centering
    \includegraphics[width=0.5\textwidth]{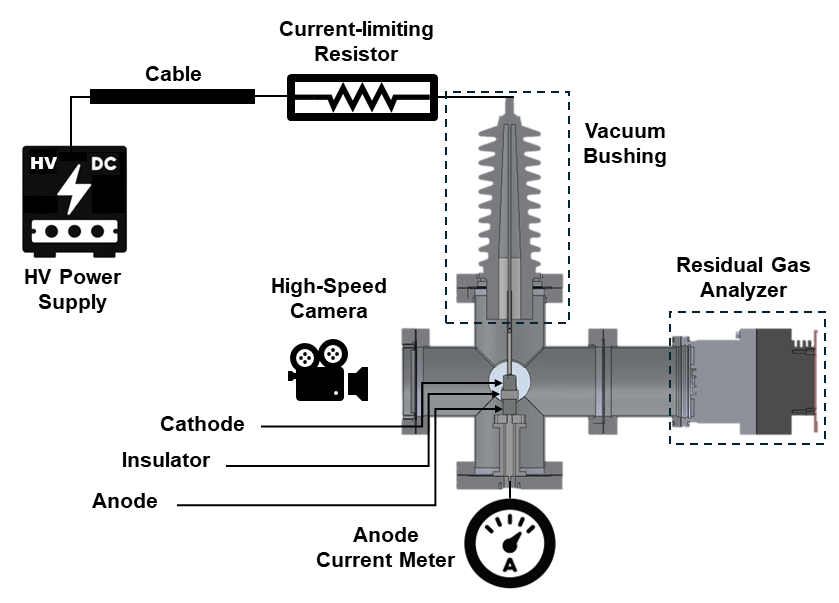}
    \caption{Schematic representation of the apparatus for the measurement of field emission current.}
    \label{fig:setup}
\end{figure}

Figure \ref{fig:setup} shows the experimental setup for field emission testing on different cathode materials. The electrode assembly involves a negatively charged cathode, a dielectric spacer (made from MACOR), and a grounded anode (fabricated from stainless steel). The assembly is housed within a vacuum chamber, where a roughing pump and a turbo pump maintain a pressure below $1\times10^{-8} \ Torr$. The high voltage is transmitted from the power supply to the cathode using a cable along with a resistor limiting current $1 \ M\Omega$. The power supply voltage and current are recorded at a rate of one sample per second. The voltage is increased incrementally to follow a current conditioning method from $0 \ kV$ to $80 \ kV$. During the experiment, measurements of the anode current, vacuum pressure, and X-ray intensity are taken along with the voltage and current read at the power supply terminal. Observations of discharges are made visually through a chamber window using a high-speed camera equipped with wide-field optics to detect any discharge inside the vacuum chamber. 

\subsection{Sample Preparation}
\begin{figure}
     \centering
     \begin{subfigure}[b]{0.35\textwidth}
         \centering
         \includegraphics[width=\textwidth]{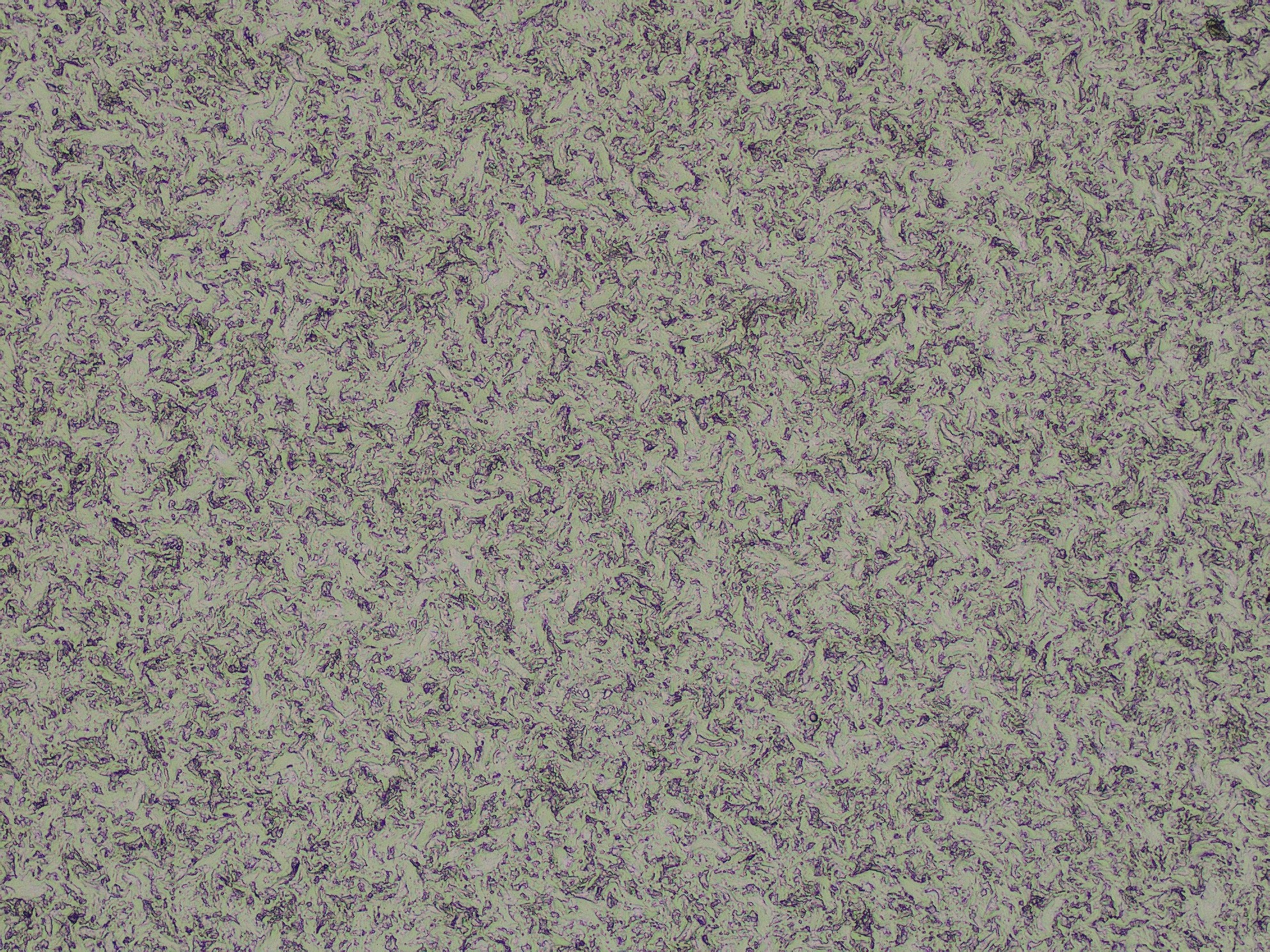}
         \caption{Polished molybdenum.}
         \label{fig:moly_10x}
     \end{subfigure}
     \hfill
     \begin{subfigure}[b]{0.35\textwidth}
         \centering
         \includegraphics[width=\textwidth]{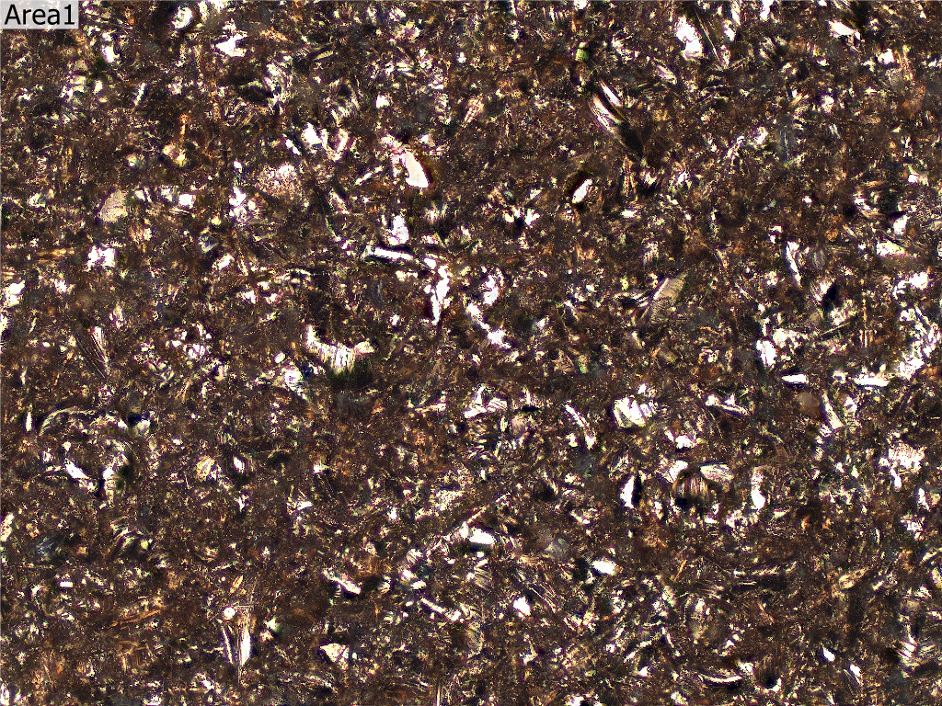}
         \caption{Grit blasted copper.}
         \label{fig:copper_10x}
     \end{subfigure}
        \caption{The surface morphology of two specimens magnified at 10x under the 3D surface profiler laser microscope.}
        \label{fig:microscope_images}
\end{figure}

All specimens for testing are made from bulk materials. The selection of cathode materials included molybdenum (type 361) with a purity of 99.95\%, titanium (grade 2), and copper (grade 101). The specimens were shaped using computer numerical control (CNC) machining. The desired surface roughness is achieved through manual sanding ($2 \ \mu m$ or finer) or grit-blasting for increased surface roughness ($4-10  \ \mu m$). Figure \ref{fig:microscope_images} illustrates the characteristic copper and molybdenum specimen under 10x magnification. 
Following fabrication, the samples are subjected to a cleaning protocol to remove hydrocarbons and other residues that remain from the manufacturing process. Initially, each sample is washed with hot soapy water in a sink to remove bulk contaminants. Next, each sample is subjected to a 15-minute ultrasonic cleaning cycle with hot soapy water. After being rinsed, each sample is then transferred to another ultrasonic cleaner containing deionized water for a brief two-minute cleansing. Subsequently, the sample is immersed in an ultrasonic cleaner filled with isopropyl alcohol (IPA) for an additional 15 minutes. The final step includes drying the sample with a lint-free cloth in a class-10,000 cleanroom.

\section{Data Preprocessing} \label{sec:data}
To prepare the raw experimental and simulation data for use in a data-driven model, the data must be cleaned and enriched with additional information. The flow chart in Figure \ref{fig:flowchart} shows the steps discussed in the rest of the section.

\subsection{Experimental Data}
%As previously stated, this study uses more than 259 hours of data. The data processing ensures that only the stable cathode current values post-conditioning are obtained for each voltage level. This means that transient variations in the cathode current following a voltage increase are excluded from the training dataset.

The first step is the removal of the data points where, due to electromagnetic interference and transient faults, the voltage or current has not been properly stored. In the next step, the tests are automatically separated by tracing the rise and fall of the cathode voltage in the dataset. Any test missing information on the cathode condition is deleted. Data points where the voltage setpoint and the voltage measurement deviate by greater than 10\% are also removed.

%Per conditioning procedure, the voltage has been increased in 2kV increments, so it is obvious that only a finite set of voltages will be included in the data set. 
The data collected from each test are acquired only once the cathode has been conditioned to a specific voltage level. The conditioning process effectively eliminates problematic emitters and adsorbates by gradually increasing the voltage until the current stabilizes. As a result, only stable field emission current values are considered, facilitating the reproducibility of the results.

The experimental data used in this model are measurements of cathode voltage, $V_{cathode}$, and cathode current, $I_{cathode}$. However, the experimental setup measures the current of the power supply, $I_{PS}$, and the voltage, $V_{PS}$. Since there is a series resistor, $R_{series}$, in line between the power supply and the vacuum chamber, the measured voltage ($V_{PS}$) includes an additional voltage due to the drop across the resistor. Therefore, the cathode voltage is calculated as $V_{cathode} = V_{PS} - R_{series} \times I_{PS}$.

Additionally, the current, $I_{PS}$, captures not only the cathode current but also any leakage current, $I_{leakage}$, such as that of the corona in the exposed electrodes. Therefore, the baseline leakage current as a function of the power supply voltage is deducted from the actual current measurement at each corresponding voltage level. By deducting the leakage current from the measured current, we acquire the cathode current $I_{cathode}=I_{PS}-I_{leakage}$. Ultimately, we obtain a matrix containing numerous pairs of $V_{cathode}$ and $I_{cathode}$, with each pair representing the total field emission current of a specific cathode under a particular voltage.

\subsection{Electrostatic Simulation}
The field emission of electrons is strongly related to the strength of the electric field on the cathode surface, as discussed in Section \ref{sec:FN}. For a single emitter, the field at that emission site is needed to calculate the emission current based on the Fowler-Nordheim expression. However, in the case of broad-area emission, the electron current stems from a multitude of emission sites. The apex (or maximum) value of the local electric field at each of these emission sites depends on (1) the macroscopic electric field and (2) the microscopic field enhancement factor. In this study, the objective is to incorporate these parameters statistically. As a first step, the macroscopic electric field at the surface of the cathode is simulated electrostatically.

\begin{figure}[H]
     \centering
     \begin{subfigure}[b]{0.45\textwidth}
         \centering
         \includegraphics[width=\textwidth]{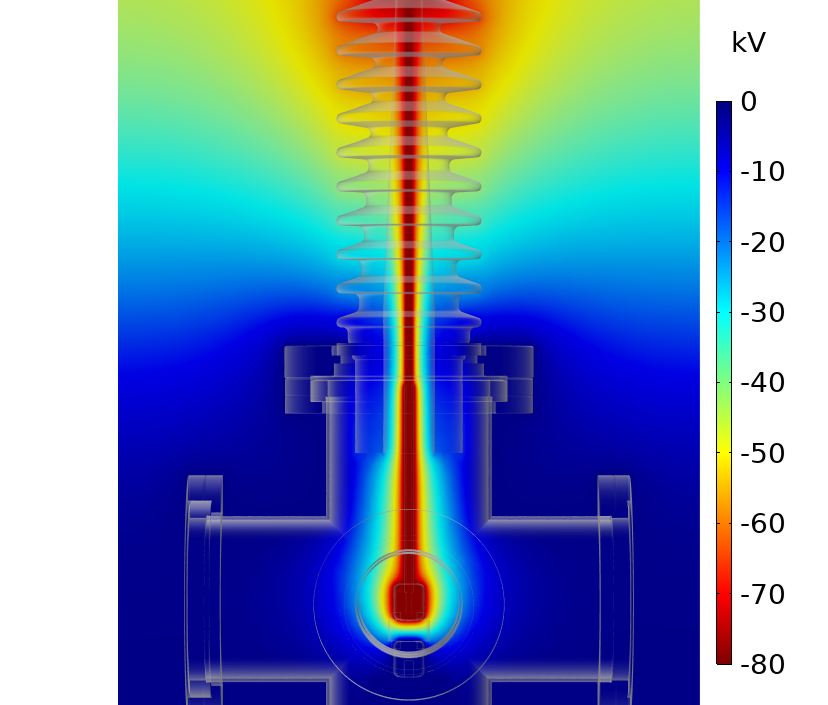}
         \caption{Electric potential distribution.}
         \label{fig:comsol_voltage}
     \end{subfigure}
     \begin{subfigure}[b]{0.45\textwidth}
         \centering
         \includegraphics[width=\textwidth]{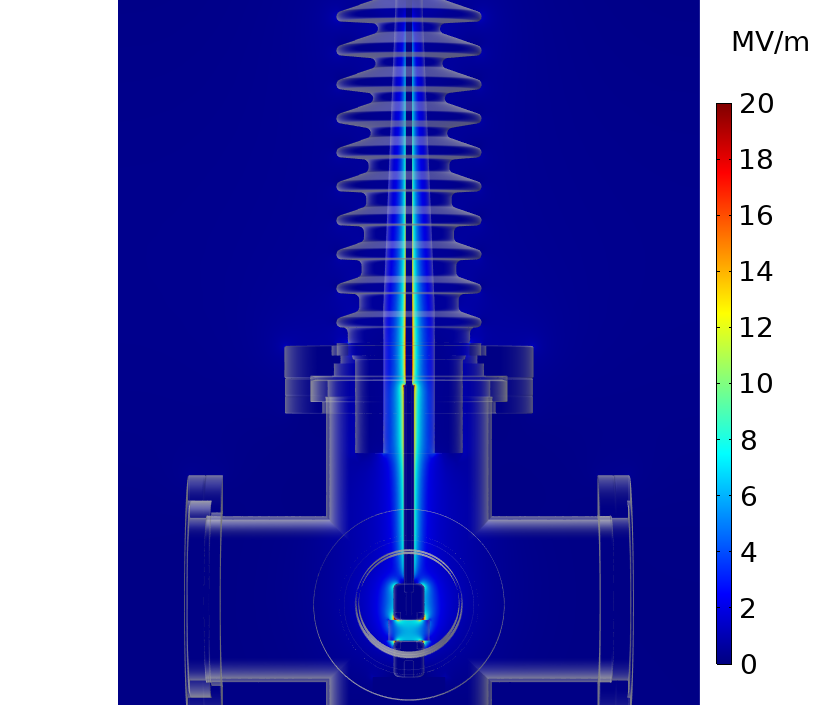}
         \caption{Electric field distribution.}
         \label{fig:comsol_field}
     \end{subfigure}
        \caption{The electrostatic simulation of the high voltage assembly with cathode at -80kV.}
        \label{fig:comsol}
\end{figure}

Figure \ref{fig:comsol} shows the distribution of the electric field in the periphery of the cathode. To feed the machine learning model, this study uses statistical features that describe the inhomogeneity of the electric field distribution. For a virgin electrode, there is a threshold field in the range of $5-15 \ MV/m$ associated with the initial switch-on of the field emission current \cite{LATHAM_1991}. Therefore, the values of the electric field below $5 \ MV/m$ are excluded from the probability distribution of the electric field.\\

A Gaussian mixture model (GMM) is used to represent an inhomogeneous electric field distribution at the surface of a cathode by assuming that the field distribution is composed of multiple Gaussian components, each representing a distinct region with its own mean and variance. This model allows for the accommodation of complex and multi-modal field distributions, which are common in practical scenarios due to irregularities in material properties and geometric configurations. Each component in the GMM can be thought of as a local 'peak' or 'valley' in the field intensity, and the overall model provides a probabilistic framework for estimating the likelihood of field strengths at different points on the cathode surface. The sensitivity analysis of model prediction with respect to the number of components are discussed in section \ref{sec:discuss}. 

\begin{figure}
    \centering
    \includegraphics[width=0.5\textwidth]{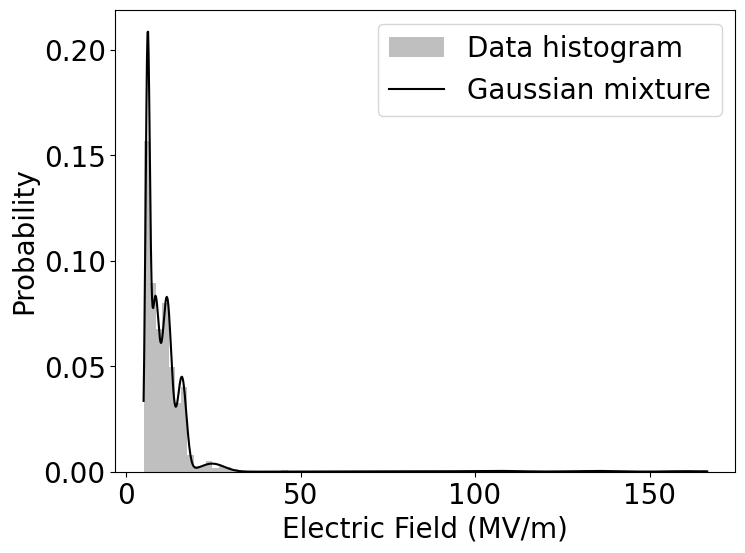}
    \caption{Probability distribution of electric field at the surface of cathode.}
    \label{fig:field_gmm}
\end{figure}

Figure \ref{fig:field_gmm} demonstrates the distribution of the electric field intensity using four Gaussian distributions. Therefore, eight statistical features are used to explain the contribution of the electric field to the field emission in the model. After benchmarking different numerical models. A sensitivity analysis of the number of components in the GMM approximation is performed to find the optimal mixture model.

\subsection{Optical Microscopy}
For broad-area electrodes, it is well-established that field enhancement can exhibit significant spatial variability. Consequently, an effective approximation of surface microprotrusions is essential to capture the extensive variance in the current-voltage measurements. In this study, we employ laser microscopy to characterize the surface profile of the sample, thereby enabling a detailed examination of the microscopic features of the cathode surface. The chemical composition has previously been analyzed by the authors in \cite{borghei_2023} and is not repeated here for brevity. 

\begin{figure}[h]
     \centering
     \begin{subfigure}[b]{0.4\textwidth}
         \centering
         \includegraphics[width=\textwidth]{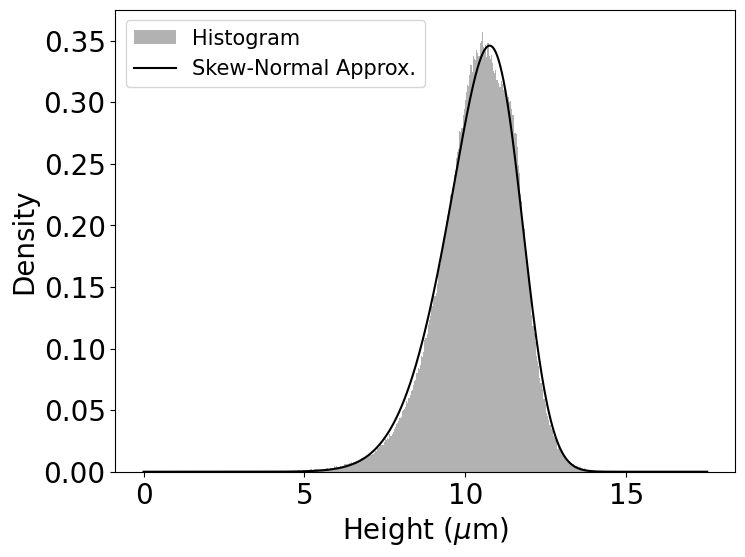}
         \caption{Polished molybdenum.}
         \label{fig:moly_dist}
     \end{subfigure}
     \vspace{5mm}
     \begin{subfigure}[b]{0.4\textwidth}
         \centering
         \includegraphics[width=\textwidth]{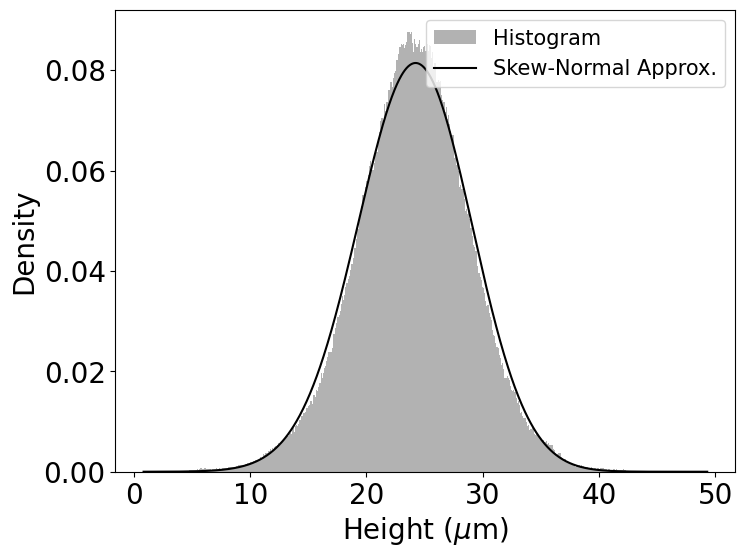}
         \caption{Grit blasted copper.}
         \label{fig:copper_dist}
     \end{subfigure}
        \caption{The probability distribution of projection heights for two specimens.}
        \label{fig:dist_heights}
\end{figure}

Surface roughness is determined with a Keyence VK X3050 3D surface profiler. Each specimen is examined using the laser microscope and the height array for an area of $1 \ \mu m \times 1 \ \mu m$ at the center of the cathode surface. Due to the uniform surface preparation process, the $1 \ \mu m^2$ area can provide a valid representation of the total cathode area. Figure \ref{fig:dist_heights} depicts the probability distributions of the surfaces of two specimens shown in Figure \ref{fig:microscope_images}. If one plots the histogram of the height distribution, a skewed Gaussian distribution can approximate the projection heights of the probability distribution. Therefore, three statistical features incorporate a microscopical fingerprint.

%mean= 10.391900915193125 , std dev= 1.2171675674280775 , data skewness= -0.675229131054281

%mean= 24.07699545968464 , std dev= 4.912981826169416 , data skewness= -0.04220136780993071

\begin{figure*}
    \begin{tikzpicture}

% Define positions for the columns with more space for E-Field parameters and reduced space between last parameter and output
\foreach \i in {1,2,...,11}
  \coordinate (A\i) at (1.25*\i-1.25,0);
\coordinate (A12) at (14.2,0); % Slightly closer to the last E-Field parameter
\coordinate (B1) at (15.5,0);  % Cathode Current

% Draw table columns and feature labels with smaller font size
\node at (A1) [align=center, font=\footnotesize] {Cathode \\ Voltage};
\node at (A2) [align=center, font=\footnotesize] {Cathode \\ Area};
\node at (A3) [align=center, font=\footnotesize] {Work \\ Function};
\node at (A4) [align=center, font=\footnotesize] {Height \\ Mean};
\node at (A5) [align=center, font=\footnotesize] {Height \\ Variance};
\node at (A6) [align=center, font=\footnotesize] {Height \\ Skewness};
\node at (A7) [align=center, font=\footnotesize] {E-Field \\ Mean \\ 1};
\node at (A8) [align=center, font=\footnotesize] {E-Field \\ Variance \\ 1};
\node at (A9) [align=center, font=\footnotesize] {...};
\node at (A10) [align=center, font=\footnotesize] {E-Field \\ Mean \\ N};
\node at (A11) [align=center, font=\footnotesize] {E-Field \\ Variance \\ N};
\node at (B1) [align=center, font=\footnotesize] {Cathode \\ Current};

% Draw lines between features
\foreach \i in {1,2,...,10} {
  \draw[dashed] (1.25*\i-0.625, -0.5) -- (1.25*\i-0.625, 0.5);
}
\draw[dashed] (13.1, -0.5) -- (13.1, 0.5); % Line before last parameter

% Draw the horizontal accolade on top
\draw[decorate,decoration={brace,amplitude=10pt,raise=4pt},thick]
(0,0.6) -- (13.1,0.6) node[midway,yshift=25pt, font=\footnotesize]{2N+6 Input Features};

\draw[decorate,decoration={brace,amplitude=10pt,raise=4pt},thick]
(15,0.6) -- (16.1,0.6) node[midway,yshift=25pt, font=\footnotesize]{Output Feature};

% Draw an arrow from the features to the output
\draw[->, thick] (13.3,0) -- (14.7,0);

\end{tikzpicture}
\caption{The structure of single datapoint in the input and output arrays.}
\label{struct}
\end{figure*}
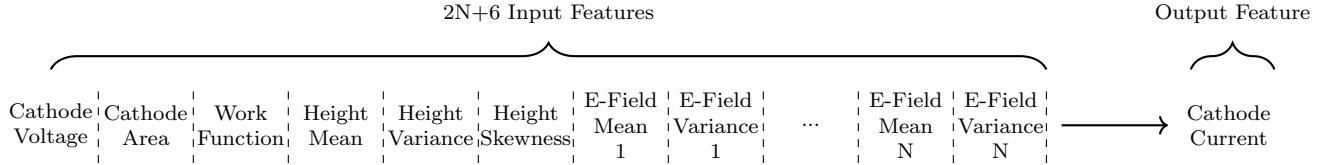

\section{Numerical Results}\label{sec:results}
After the data have been processed and the datasets have been created, a collection of physical and environmental parameters is used as input features, and the cathode current vector serves as the target vector. The layout of a single row of input/output features can be seen in Figure \ref{struct}. Only stable field emission current data points post-conditioning are utilized in the modeling. The total count of data points is 920, which will be divided into 80\% for training and 20\% for testing. This section benchmarks several well-known models. Given the widespread use of these models in different areas of machine learning, descriptions of each algorithm are avoided (for comprehensive information on these models, refer to \cite{bishop2006pattern}).

The hyperparameters of each model can be adjusted to improve the accuracy of the prediction. We begin with simple models such as linear regression, and then move toward more advanced models. Performance metrics used include: (1) mean absolute error (MAE) that calculates the average absolute difference, treating all errors equally, (2) mean squared error (MSE), which measures the average squared difference between predicted and actual values, penalizing larger errors more severely, and (3) the coefficient of determination ($R^2$ score) that quantifies the proportion of variance in the dependent variable explained by the model, with 1 indicating perfect prediction and 0 indicating no explanatory power. To find the optimal hyperparameters, we utilize the Bayesian optimization method, specifically the tree-structured Parzen estimator, with the aim of maximizing the $R^2$ score. The models are developed in Python using open-source libraries, including \verb+Pandas+, \verb+NumPy+, \verb+TensorFlow+, \verb+Scikit-learn+, \verb+Optuna+, and \verb+XGBoost+.

\subsection{Linear Regression} 

Linear regression is the simplest regression model if a linear correlation is anticipated between the input and the target variable. In the first attempt, we have employed Ridge and Lasso in linear regression as regularization strategies to attenuate overfitting. 
%which can be effective where multicollinearity exists among predictors 

The focus of hyperparameter tuning is on the regularization parameter \(\alpha\), exploring a logarithmic scale from \(10^{-4}\) to \(10^{4}\). Through 1000 trials, the optimal \(\alpha\) values for Ridge and Lasso were found to be approximately 26.58 and 1.07, respectively. With these optimal parameters, the models were trained on the training dataset and then evaluated on the test dataset. The Ridge model achieves an MSE of 1414.1, an MAE of 26.9, and an R\textsuperscript{2} score of 43.8\%. In contrast, the Lasso model achieves an MSE of 1529.8, an MAE of 27.4, and an R\textsuperscript{2} score of 39.2\%. The R\textsuperscript{2} score indicates that the Ridge regression model explains approximately 43. 8\% and 39. 2\% of the variance in the target variable, respectively. The performance of these models indicates a necessity for more sophisticated modeling techniques to better capture the highly nonlinear relationship between field emission current and its input features, thereby enhancing predictive accuracy.

\subsection{Support Vector Regression}

Support Vector Regression (SVR) has shown to be effective in high-dimensional spaces and when a certain tolerance margin within the model is required. It often achieves good performance with non-linear relationships when the suitable kernel is employed. In our implementation of SVR, hyperparameter optimization is carried out focusing on the regularization parameter, \(\alpha\), which ranged from \(10^{-4}\) to \(10^{4}\), \(\epsilon\) spanning from \(10^{-4}\) to \(10^{2}\) defining the tolerance margin where no penalties are applied for errors, kernel types including linear, polynomial, radial basis function (RBF), and sigmoid. From more than 1000 trials, the optimal parameters determined were $\alpha \ = \ 4967.8$, $\epsilon \ = \ 0.27$, RBF kernel, and its tuning parameter $\gamma \ = \ 3.06$, which describes the extent of influence of a single training example.
%, as well as the gamma parameter, $\gamma$, from $10^{-6}$ to $10^3$, 

Using these optimal parameters, the SVR model is trained on the full training dataset and evaluated on the test dataset. The SVR model achieves an MSE of $1051.67$, an MAE of $18.67$, and an R\textsuperscript{2} score of 58.2\%. The improvement in predictive accuracy, compared to the Ridge and Lasso regression models, highlights the higher efficacy of SVR in capturing nonlinear relationships within this set of data. Nonetheless, the results suggest potential for further enhancements in predictive performance.
%This R\textsuperscript{2} score indicates that the SVR model explains approximately 58.\% of the variance in the target variable. 

\subsection{Random Forest Regression}

In the next model, we explore random forest regression (RFR), an ensemble learning method that builds multiple decision trees and combines them to achieve more accurate and stable predictions. This approach effectively handles complex interactions and nonlinearities, providing better generalization and feature importance scores, though it is less interpretable and computationally intensive. The key optimization parameters included the number of estimators (from 100 to 10000), maximum tree depth (from 10 to 1000), minimum samples for splitting (from 2 to 20) and leaf nodes (from 1 to 20), maximum number of features for splitting (from 4 to 14) and bootstrap sampling. Through 1000 optimization trials, the optimal parameters identified were 7965 estimators, a maximum depth of 316, a minimum of 2 samples for node splitting, 2 samples per leaf node, the maximum number of features of 14 without bootstrapping. 

The RFR model achieved an MSE of $82.6$, an MAE of $4.4$, and an R\textsuperscript{2} score of $96.7\%$. This high R\textsuperscript{2} score indicates that the random forest model accounts for approximately 97\% of the variance in the target variable, demonstrating its superior performance in terms of capturing complex relationships within the dataset compared to previous models.

\subsection{Gradient Boosting Regression}

Further extending our analysis, we applied gradient boost regression (GBR), another ensemble technique that builds models sequentially to focus on the errors of previous models, often enhancing accuracy. GBRs are effective in achieving high predictive accuracy, but can be prone to overfitting, sensitive to noisy data and outliers, and require meticulous parameter tuning. The optimization process involved parameters such as the number of estimators (100 to 10,000), learning rate (0.01 to 0.5), maximum depth (1 to 10), minimum samples to split a node (2 to 20), minimum samples in a leaf node (1 to 20), subsample ratio (0.25 to 1.0), and the maximum features to split. After 1000 trials, the optimal parameters identified were 7895 estimators, a learning rate of 0.43, a maximum depth of 2, 2 samples to split a node, 10 samples per leaf node, a subsample ratio of 0.985, and the maximum number of features of 13.

The GBR model achieved an MSE of 42.1, an MAE of 3.2, and an R\textsuperscript{2} score of $98.2\%$. This high R\textsuperscript{2} score indicates that the gradient boost model accounts for more than 98\% of the variance in the target variable. The significant reduction in error metrics and the impressive R\textsuperscript{2} score underscore the robustness of gradient boosting to capture complex data patterns and relationships.

\subsection{Extreme Gradient Boosting Regression}

Compared to GBR, Extreme Gradient Boosting Regression (XGBR) uses more advanced regularization techniques, handles missing values more efficiently, provides parallel and distributed computing, and offers more options for customization. The parameters considered include the column subsampling ratio (varying from 0.1 to 1.0), which determines the proportion of features used for each tree, the learning rate (from 0.01 to 0.3), the maximum depth of the trees (from 3 to 50), the $L_1$ regularization term on weights, $\alpha$ (varying from 0.1 to 100), the $L_2$ regularization term on weights, $\lambda$ (from \(10^{-8}\) to 1), and the number of boosting rounds (from 100 to 1000). Additional parameters include the minimum loss reduction required for the partition of leaf nodes, $\gamma$ (from \(10^{-8}\) to 1), the minimum weight required to create a new tree node (from 1 to 10), and the controlling factor for the balance of positive and negative weights (between 0.1 and 10).

The best tuning results for the XGBR model indicate that the optimal parameters include a tree feature selection rate of approximately 36\%, a learning rate of about 29.4\%, and a maximum tree depth of 29. Regularization is achieved with an $\alpha$ value of approximately 0.15 and a $\lambda$ value near 0.002, while the model is trained using 604 estimators. The optimal model also has a $\gamma$ value of around 0.003, a sample rate of approximately 64\%, and a minimum child weight of 1. Additionally, the scale for positive class weights is set at about 7.56. The best model achieves an $R^2$ score of 90.9\%, an MSE of 229.0 and an MAE of 7.5.
The results demonstrate a relatively robust performance of the optimal XGBR model, but inferior to GBR and RF, due to the ineffectiveness of the extra regularization parameters given the size of the dataset and the number of features.

\subsection{Neural Network Regression}

Neural networks are highly flexible and powerful, capable of learning very complex patterns and relationships in large and complex datasets. They offer significant flexibility, but require careful tuning, are prone to overfitting, and generally lack interpretability. More data and computational resources might be needed.

The neural network architecture is optimized by fine-tuning several crucial parameters. These include the number of hidden layers (varying from 1 to 8) and the number of units per layer (ranging from 16 to 128), the activation function (chosen from linear, ramp, softmax, sigmoid, hyperbolic tangent, exponential and scaled exponential linear units.), the dropout rate (fluctuated between 0.0 and 0.5), the learning rate ($\eta$) from $1e^{-5}$ to $1e^{-1}$, and the optimizer including various versions of adaptive moment estimation (Adam, Adamax, NAdam), root mean square propagation (RMSProp) and gradient-based optimization methods (SGD, AdaDelta, AdaGrad, FTRL). In addition, the number of epochs can be up to 200 with early stopping on patience with 5 epochs, and the batch size is adjusted between 32 and 64.

The optimal hyperparameters identified were as follows: a network architecture consisting of 2 layers (first with 32 neurons and 12\% dropout, and second with 128 neurons and 31\% dropout), ramp activation function, RMSProp optimizer, a 2.4\% learning rate and a batch size of 32. The best result from the neural network model achieves MSE, MAE and $R^2$ values of 922.9, 20.3 and 63.4\%, respectively. 
Although this performance surpasses that of linear models, it does not match the effectiveness of ensemble methods. The low number of layers in the optimal shows the need for a larger dataset and possibly a higher number of features to optimize their performance.

\begin{comment}

\begin{table*}
\caption{\label{ml-models}Comparison of Machine Learning Models}
\begin{tabular*}{\textwidth}{@{}l*{3}{@{\extracolsep{\fill}}p{4cm}}}
\br
Model & Type & Pros & Cons \\
\mr
Linear Regression & Linear Model & Simple, Interpretable, Quick to Train & Assumes linearity, not suitable for complex relationships \\
\mr
Ridge/Lasso Regression & Linear Model with Regularization & Handles multicollinearity, Prevents overfitting & Assumes linearity, Regularization strength needs tuning \\
\mr
Decision Tree Regression & Tree-Based Model & Captures non-linear relationships, Easy to interpret & Prone to overfitting, Sensitive to small data changes \\
\mr
Random Forest Regression & Ensemble Model & Reduces overfitting, Handles high dimensionality & Less interpretable, Computationally intensive \\
\mr
Gradient Boosting Machines & Ensemble Model & High accuracy, Handles complex data well & Prone to overfitting, Sensitive to hyperparameters \\
\mr
Support Vector Regression & Kernel-Based Model & Effective in high-dimensional spaces, Robust to overfitting & Requires parameter tuning, Computationally intensive \\
\mr
Neural Networks & Neural Model & Highly flexible, Captures complex relationships & Requires large data, Prone to overfitting, Complex to tune \\
\br
\end{tabular*}
\end{table*} %}
\end{comment} 

\begin{figure}
    \centering
    \includegraphics[width=0.5\textwidth]{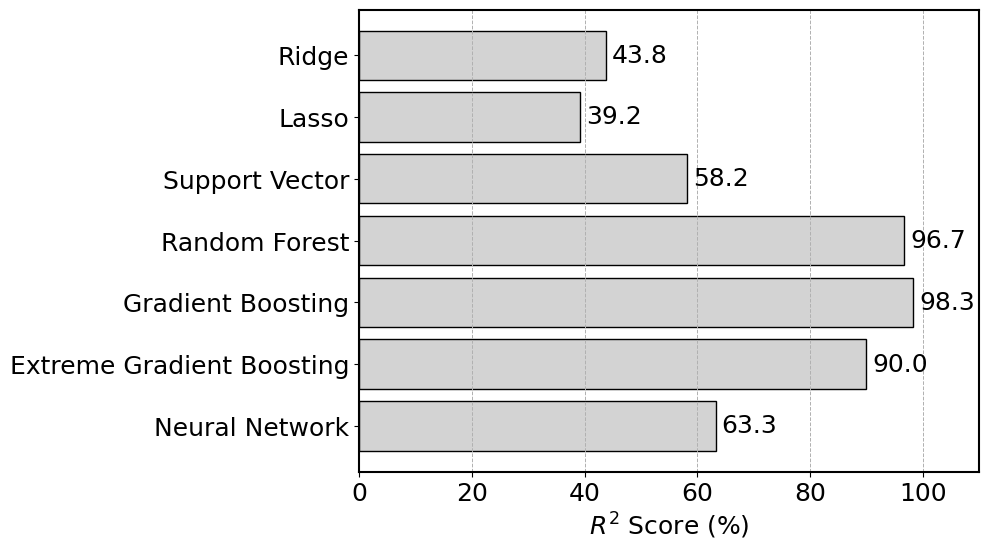}
    \caption{Performance comparison among various machine learning models.}
    \label{fig:perf_comp}
\end{figure}

\section{Discussion}\label{sec:discuss}
The performance of various regression models on the given dataset, indicated by their $R^2$ scores, is summarized in Figure \ref{fig:perf_comp}. The results reveal that traditional linear models such as Ridge and Lasso exhibit relatively low $R^2$ scores of 43.8\% and 39.2\%, respectively, suggesting a limited ability to capture complex relationships in the data. In contrast, Support Vector Regression achieves a slightly higher score of 58.2\%, indicating its ability to handle nonlinear patterns marginally better than Ridge and Lasso. However, it is the ensemble methods that significantly outperform the others: random forest, gradient boosting, and extreme gradient boosting show $R^2$ scores of 96.7\%, 98.3\%, and 90.0\%, respectively. These high scores highlight the strength of ensemble learning in reducing variance and bias through the aggregation of multiple models. It is of note that the performance of the neural network has been subpar compared to the ensemble models, with an $R^2$ score of 63.3\%.

The superior performance of ensemble methods such as random forest and gradient boosting can be attributed to their inherent ability to combine the strengths of multiple decision trees, thereby enhancing the overall robustness and accuracy of the model. Random forest achieves high performance through bagging, which reduces overfitting by averaging multiple decision trees trained on different subsets of the data. Gradient boosting, on the other hand, builds trees sequentially, each correcting the errors of its predecessor, leading to a highly accurate model, as evidenced by its highest $R^2$ score of 98.3. Extreme gradient boosting, a more efficient and regularized version of gradient boosting, also performs exceptionally well with a score of 90.0, balancing computational efficiency and accuracy. Neural networks, with their deep learning capabilities, perform better than traditional linear models but fall short of ensemble methods, likely due to their need for extensive tuning and larger datasets to achieve peak performance. The variation in these scores underscores the importance of model selection based on the specific characteristics and complexity of the dataset at hand.

A sensitivity analysis on the effect of the GMM components (1 to 9) on the electric field approximation is also performed to further enhance the accuracy of the GBR prediction. However, the $R^2$ score remains between 97.0\% and 98.6\% without any trend, indicating minimal impact from varying electric field approximation complexity. This is somewhat expected because of the identical cathode geometry in all data points. The impact is expected to be more significant with the varied cathode geometries.
%Throughout this iterative process, we observed several impactful factors:
%1. **Model Complexity and Diversity**: Ensemble models, particularly those that combined diverse algorithms, consistently outperformed single models. The inclusion of models with different learning biases helped capture various aspects of the data.
%2. **Feature Engineering**: Generating polynomial features and interactions significantly improved model performance by providing additional insights into the relationships between variables.
%3. **Hyperparameter Optimization**: Tools like Optuna were instrumental in fine-tuning model parameters, thereby enhancing predictive accuracy.
%4. **Weighted Blending**: Assigning weights based on individual model performance ensured that the ensemble benefited from the strengths of the best models while mitigating the impact of weaker ones.

\section{Conclusion}\label{sec:conclude}
In pursuit of a robust yet practical model to predict total field emission from broad-area electrodes, this study approached the problem with an emphasis on extensive experimental data. We paired the experimental data with the electrostatic field distribution, surface profile, and material information to build a matrix of input features for prediction. Upon benchmarking various machine learning models, we found that the ensemble models, especially the gradient boosting regression, perform remarkably well, achieving over 98\% accuracy. Neural neural networks of various depths did not offer improved results.

The findings presented herein should not be considered definitive. As the number of features and/or the size of the dataset expands, more sophisticated models may become necessary. These advanced models will potentially encapsulate a broader range of variability and, it is anticipated, will eventually evolve into a sufficiently robust framework capable of accurately predicting the field emission current from an electrode.
One should also note that more statistical features may hinder the practicality and cost of using such models.

Future studies will be directed towards utilizing more extensive datasets that cover a variety of cathode geometries and involve multiple work functions for each material corresponding to different crystal faces. Additionally, integrating additional physical phenomena into the simulation models can broaden the statistical distribution related to the field emission of unconditioned cathodes.
%There are instances such as particle accelerators, high voltage switches, capacitors, transmission valves, fusion reactors, traveling wave tubes, and [to be added]. In all of these cases, it is essential that none of the ‘free’ electrons leak through the electrode surface to become truly free, since in this state they are capable of initiating a variety of destructive interactions that usually result in an 'electric breakdown'.
 \section*{Acknowledgment}
We express our gratitude to Alexander Fiore, Madeline Vorenkamp, Craig Miller and Rod Latham for their insightful discussions and thorough review of our manuscript.

 \section*{Data Availability}
The data that support the findings of this study are available from the corresponding authors upon reasonable request.
%\section*{References}

%% If you have bibdatabase file and want bibtex to generate the
%% bibitems, please use
%%
%\bibliographystyle{IEEEtran} 

\bibliography{main}

\end{document}